\documentclass[letterpaper,notitlepage,nofootinbib,superscriptaddress]{revtex4-1}

\usepackage[utf8]{inputenc}
\usepackage{amsmath,amssymb,amsfonts}
\usepackage{graphicx,wrapfig}
\usepackage{dcolumn}
\usepackage{bm}
\usepackage{color}
\usepackage{bbold}
\usepackage{url}
\usepackage{slashed}
\usepackage[svgnames]{xcolor}
\usepackage[english]{babel}
\usepackage{microtype}

\usepackage[colorlinks=true,linkcolor=blue,urlcolor=blue,citecolor=blue]{hyperref}

\usepackage{color}

\definecolor{darkgreen}{rgb}{0,0.65,0}

\newcommand{\be}{\begin{eqnarray}}
\newcommand{\ee}{\end{eqnarray}}
\newcommand{\ba}{\begin{array}}
\newcommand{\ea}{\end{array}}

\begin{document}

\title{ Large violation of the flavour SU(3) symmetry in $\eta$MAID2018 isobar model\\
{\normalsize (the neutron anomaly case)}}

\author{V. A. Kuznetsov}
	\affiliation{Petersburg Nuclear Physics Institute,
		Gatchina, 188300, St.~Petersburg, Russia}
\affiliation{INFN - Sezione di Catania, via Santa Sofia 64, I-95123 Catania, Italy}

\author{M. V. Polyakov}
	\affiliation{Petersburg Nuclear Physics Institute,
		Gatchina, 188300, St.~Petersburg, Russia}
	\affiliation{Institut f\"ur Theoretische Physik II,
		Ruhr-Universit\"at Bochum, D-44780 Bochum, Germany}

\begin{abstract}
We demonstrate that the explanation of the neutron anomaly around $W\sim 1685$~MeV in $\gamma N\to \eta N$ reactions 
provided by the $\eta$MAID2018 isobar model  is based on  large violation of the flavour SU(3) symmetry in hadron interactions.   
This is yet another example of how conventional explanation (without invoking exotic narrow nucleon resonance) of the neutron anomaly  metamorphoses into  unconventional 
physics picture of hadron interactions.

A possibility to mend the flavour SU(3) symmetry for some of resonances in $\eta$MAID model is discussed.

\end{abstract}


\maketitle


\section{ Introduction}

\begin{figure}[h!]

\vspace{3mm}

\centering
\includegraphics[height=7cm]{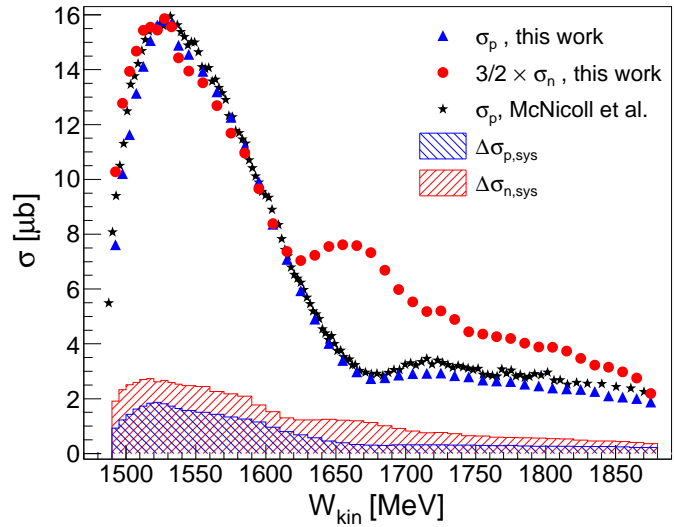}
\vspace{-1mm}
\caption{\small Figure from the third of Refs.~\cite{wert}. Total cross sections as a function of the final-state invariant mass 
$m(\eta N)$: Blue triangles: proton data. Red circles: neutron data scaled by $3/2$. 
Black stars: free proton data from MAMI-C \cite{Mainz}.
Hatched areas: total systematic uncertainties of proton (blue) and neutron (red) data.}
    \label{fig:krusche} \vspace{0.3cm}

\end{figure}

The discovery of the neutron anomaly\footnote{ Existence of the narrow ($\Gamma\sim$10-40~MeV) peak in the $\gamma n\to \eta n$ cross section around $W\sim 1685$~MeV and small dip 
in the $\gamma p\to \eta p$ process at the same invariant energy.} in  the $\gamma N\to\eta N$  cross section was reported first in Ref.~\cite{gra0}.
Presently, three other experiments 
( LNS~\cite{kas},CBELSA/TAPS\cite{kru}, and A2~\cite{wert}) confirmed the neutron anomaly beyond any doubts.  
For an illustration of the phenomenon, we show   the recent results of the A2 collaboration 
 on Fig.~\ref{fig:krusche}.
Furthermore the neutron anomaly at the same invariant mass of $W\sim 1685$~MeV 
was also observed in the Compton scattering \cite{Compton}. More recently the corresponding narrow structures  were observed in
spin asymmetry $\Sigma$ for the Compton scattering off the proton \cite{Kuznetsov:2015nla} and in $\gamma N \to \pi\eta N$ processes \cite{Kuznetsov:2017xgu}. 
The precise $\pi N$ scattering data also provided an evidence for the narrow structures \cite{Gridnev:2016dba}.

Following the {\it Prinzip der Denk\"okonomie} the simplest and universal (in all channels) explanation of the neutron anomaly would be the assumption of
new narrow nucleon resonance with stronger photo-coupling to the neutron. This option was suggested before the discovery of the neutron anomaly \cite{max} (see Appendix
for a reminder about this option). There are several conventional  (without invoking exotic narrow nucleon resonance) explanations of the neutron anomaly, for critical discussion of these explanations
see e.g.  Refs.~\cite{Boika:2014aha,Kuznetsov:2017ayk}. 

\section{Flavour SU(3) symmetry constraints versus $\eta$MAID2018}

In this notes we concentrate on yet another conventional explanation of the neutron anomaly in the $\eta$MAID2018 isobar model \cite{Tiator:2018heh}.
The $\eta$MAID isobar model provides a fitting function for the photo-production amplitudes with a number of free parameters. The later are adjusted to describe
the current values of  experimental observables. In recent fit \cite{Tiator:2018heh} the parameters were adjusted  to describe  the
peculiarities of the cross section of $\gamma N\to \eta N$ reactions with good $\chi^2$.   
Concerning the neutron anomaly the authors of \cite{Tiator:2018heh} stated:
\\

\noindent
{\it `` Our analysis shows that the narrow bump in $\eta n$ and the dip in $\eta p$ channels have different origin. The first is a result of an interference of 
few resonances with a dominant contribution of the $P_{11}(1710)$. The second one is mainly a sum of $S_{11}(1535)$ and $S_{11}(1650)$ with opposite signs. 
However the narrowness of this structure is explained by a cusp effect due to the opening of the $K\Sigma$ decay channel of the $S_{11}(1650)$ resonance."}\\

\noindent
Let us inspect the values of  resonance parameters needed in the fit of Ref.~\cite{Tiator:2018heh} to explain the neutron anomaly. The values of the parameters
are summarise in Table~\ref{Table:parameters}.

\renewcommand{\arraystretch}{1.3}
\begin{table}[h]
\begin{center}%
\begin{tabular}
[c]{|l|c|c|c|c|c|c|}\hline
  &  $M_R$~(MeV)&$\Gamma_R$~(MeV)  & $|g_{\pi N}|$ & $|g_{\eta N}|$ & $|g_{K \Lambda}|$ & $|g_{K \Sigma}|$\\\hline
$N\left(1535,\frac12^-\right)$    &  $1521.7 $ &$174.7$  & $0.45$ & $0.62$ & $N/A$ & $N/A$\\\hline
$N\left(1650,\frac12^-\right)$    &  $1626.3 $ &$132.5$  & $0.36$ & $0.28$ & $0.20$ & $1.21$\\\hline
$N\left(1710,\frac12^+\right)$    &  $1669.5 $ &$63.2$  & $0.10$ & $0.22$ & $0.52$ & $0$\\\hline
  \end{tabular}
\end{center}
\par
\vspace{-0.3cm} \vspace{0.3cm}\caption{ Resonance parameters used Ref.~\cite{Tiator:2018heh} to describe the neutron anomaly. 
See Table~IV and Eq.~(50) of
Ref.~\cite{Tiator:2018heh}.
The absolute values of the coupling constants are given. $N/A$ marks undetermined in fit of Ref.~\cite{Tiator:2018heh} couplings  }
\label{Table:parameters}
\end{table}
\renewcommand{\arraystretch}{1}
\noindent
The approximate flavour SU(3) symmetry of strong interactions implies that the octet resonance coupling constants are related to each other by \cite{SU3}:

\be
\label{Eq:SU3relations}
\frac{g_{\eta N}}{g_{\pi N}}=\frac{1}{3} \left(4\alpha-1\right), \quad \frac{g_{K\Lambda}}{g_{\pi N}}=-\frac{1}{3} \left(2\alpha+1\right),\quad \frac{g_{K\Sigma}}{g_{\pi N}}= \left(2\alpha-1\right).
\ee
Here $\alpha$ is related to the $F/D$ ratio as $\alpha=F/(D+F)$. In QCD, the accuracy of  the formulae above is the order of
$O\left({m_s}/{M_{\rm strong}}\right)$ with $m_s$-strange quark mass and $M_{\rm strong}\sim 1$~GeV is a typical scale in the problem. Detailed studies of the flavour 
SU(3) symmetry for resonance couplings were performed in Refs.~\cite{SU3,SU31}. It was demonstrated that the SU(3) relations are satisfied with typical accuracy
of $15-20\%$ as it  is expected for  $O\left({m_s}/{M_{\rm strong}}\right)$.\\

\noindent
Now we confront the SU(3) relations (\ref{Eq:SU3relations}) with the $\eta$MAID2018 results (see Table~\ref{Table:parameters}):

\begin{itemize}
\item
for $N\left(1535,\frac12^-\right)$ resonance only two couplings are available, so we can determine the value of $\alpha$ for this resonance. Two possible
values of $\alpha=1.28$ and $\alpha=-0.78$ can be obtained. The second value is in agreement with the value of $\alpha=-0.53\pm0.15$ obtained in the global SU(3) analysis of
baryon octets  \cite{SU31},
as well as with the SU(6) quark model value of $\alpha_{\rm SU(6)}=-1/2$. 

\item
for $N\left(1650,\frac12^-\right)$ resonance it is impossible to describe $\eta$MAID2018 couplings with SU(3) relations (\ref{Eq:SU3relations}).  Furthermore, we can use the values of
$|g_{\pi N}|$ and $|g_{\eta N}|$ from Table~\ref{Table:parameters} to obtain the value of $\alpha$ from SU(3) relations.  One obtains two possibilities $\alpha=0.83$ and $\alpha=-0.33$.
With these values, we can compute $g_{K\Lambda}$ and $g_{K\Sigma}$ couplings with the result $|g_{K\Lambda}|=0.32$, $|g_{K\Sigma}|=0.24$ (for $\alpha=0.83$), and
$|g_{K\Lambda}|=0.04$, $|g_{K\Sigma}|=0.60$ (for $\alpha=-0.33$). Both sets of values are in {\it acute} contradiction with the values needed by $\eta$MAID2018 fit to explain
the neutron anomaly. Especially it concerns the $g_{K\Sigma}$ coupling, crucial for $\eta$MAID interpretation of the small dip around $W\sim 1695$~MeV in the proton channel --
with the coupling satisfying the SU(3) relations the $\eta$MAID interpretation  fades away.

\item
for $N\left(1710,\frac12^+\right)$ resonance there is no way to describe $\eta$MAID2018 couplings with SU(3) relations (\ref{Eq:SU3relations}) as well.
From $\eta$MAID2018 values of $|g_{\pi N}|$ and $|g_{\eta N}|$ we can estimate the couplings to strange particles required by the SU(3) symmetry.
There are two solutions $|g_{K\Lambda}|=0.16$, $|g_{K\Sigma}|=0.28$ (with $\alpha=1.9$), and $|g_{K\Lambda}|=0.06$, $|g_{K\Sigma}|=0.38$ (with $\alpha=-1.40$).
Again, these values are in {\it acute} contradiction with $\eta$MAID2018 values.  We also note that the corresponding values of $\alpha$ are in strong contradiction with 
$\alpha=0.32\pm 0.03$ provided by the SU(3)
analysis \cite{SU31} of the octet containing $N(1710)$ state, as well as with all known results in variants of  quark models for $N(1710)$.

\end{itemize}
Above we considered in details only the key resonances relevant for the neutron anomaly. We checked  that for almost all other resonances their couplings used by $\eta$MAID2018 also do not satisfy 
the  SU(3) symmetry constraints. Additionally, we tried different relations between the coupling constants and the branching ratios --instead of  Eq.~(50) of Ref.~\cite{Tiator:2018heh} we used Eqs.~(4,5) 
 of \cite{Polyakov:2018mow} as well. The result remains unchanged.

We can conclude  that the values of resonance parameters
used in $\eta$MAID2018 to fit the neutron anomaly are not acceptable, unless one considerably revises the common knowledge of how SU(3) works in hadron interactions or one
allows the presence of strong non-octet components in the amplitude. 

\section{Discussion}
In general, if one has a fit function with large number of free parameters, one is almost certainly able to describe fine peculiarities of observables (like in the case of the neutron anomaly)
 with reasonable $\chi^2$.\footnote{
Here we can recall famous  J.~von Neumann's joke quoted by F.~Dyson in Ref.~\cite{Dyson}.} However, in the case of the neutral anomaly such approach often
\begin{itemize}
\item
requires unnaturally fine tuning of relevant parameters,
\item
frequently the resulting parameters are either unphysical or violate fundamental symmetries. 
\end{itemize}
Several years ago in Ref.~\cite{Boika:2014aha} we illustrated this on example of the Bonn-Gatchina analysis of the neutron anomaly \cite{ani1}.
In  \cite{Boika:2014aha} we showed that the fine tuning of photocouplings for known 
$N(1535,1/2^-)$ and $N(1650,1/2^-)$ resonances  in Refs.~\cite{ani1} to describe the neutron anomaly  unavoidably leads to the picture in which the well-known $N(1650,1/2^-)$ resonance must be 
a cryptoexotic pentaquark with large $s\bar s$ Fock component. Very unconventional picture of this well studied resonance. 
In addition to this problem,
it seems that the flavour SU(3) symmetry also is strongly violated in the recent Bonn-Gatchina \cite{ani1,Anisovich:2017afs}   fit of the neutron anomaly. 
We shall discuss this in details elsewhere \cite{inprep}. 

In this notes we gave yet another example of such situation -- the conventional explanation (without invoking narrow resonance) of the neutron anomaly in the $\eta$MAID2018 isobar model
requires unconventionally large violation of the flavour SU(3) symmetry in hadron interactions. It seems, that if one uses the resonance couplings which satisfy the octet symmetry constraints,
the $\eta$MAID2018 description of the neutron anomaly fades away. 

However, this can be mended if to note interesting coincidence -- properties of the $N(1710,1/2^+)$ resonance in $\eta$MAID resemble very much the properties of the anti-decuplet nucleon.
  The photo-couplings are tuned in $\eta$MAID
such that $|A_{1/2}^p/A_{1/2}^n|\simeq 0.13$ for $N(1710)$. Exactly what is expected for the anti-decuplet nucleon (see the Appendix)! Also the total width of $N(1710)$ in $\eta$MAID 
is about two times smaller than the PDG estimate. In a sense the properties of $N(1710)$ in $\eta$MAID model are much closer to properties of anti-decuplet nucleon, than to properties listed in PDG and
to properties of a 3-quark nucleon excitation. 
It might be that 
$N(1710)$ in the $\eta$MAID isobar model actually corresponds to two states: conventional octet $N(1710)$ and a narrow anti-decuplet nucleon. 
Such hypothesis can explain strong deviation of the $N(1710)$ couplings from the octet SU(3) symmetry constraints, because under this hypothesis a new anti-decuplet representation is involved
and the octet SU(3) constraints  (\ref{Eq:SU3relations}) are modified. 

In the $\eta$MAID model the small dip in the proton channel is due to extraordinarily large coupling of the $N(1650,1/2^-)$ to $K\Sigma$, which violates strongly the flavour SU(3) symmetry.
The narrow antidecuplet $J^P=1/2^+$ resonance can also easily solve this problem -- its presence can reproduce the proton dip, see e.g. Ref.~\cite{KPT}.  

The possibility of two component nature of $N(1710)$ state was suggested in the past in Ref.~\cite{Batinic:1995kr}. 
Would be interesting to
test such possibility again with new data. 

\section*{Acknowledgement}
The work of MVP is supported by the BMBF grant 05P2018.

\section*{Appendix: Narrow nucleon from exotic anti-decuplet}

The simplest and economical physics explanation of the neutron anomaly is the existence of a narrow anti-decuplet of baryons. The existence of such narrow exotic baryon multiplet
was predicted in Ref.~\cite{dia}. 
Main properties of N$^*$ from the anti-deculpet which were predicted theoretically in years 1997-2004 (before the discovery of the neutron anomaly) are the following:
\vspace{-0.2cm}
\begin{itemize}
\item quantum numbers are $P_{11}$ ($J^P=\frac 12^+$, isospin=$\frac 12$) \cite{dia},
\vspace{-0.2cm}
\item narrow width of $\Gamma\le 40$~MeV \cite{dia,arndt,michal},
\vspace{-0.2cm}
\item mass of $M\sim 1650-1720$~MeV \cite{arndt,michal,dia1},
\vspace{-0.2cm}
\item strong suppression of  the proton photocoupling relative to the neutron one \cite{max} ,
\vspace{-0.2cm}
\item the $\pi N$ coupling  is suppressed, N$^*$  prefers to decay into $\eta N$, $K\Lambda$ and $\pi \Delta$ \cite{dia,arndt,michal}.
\end{itemize}

Detailed account for predictions and evidences for narrow anti-decuplet nucleon 
were presented at length previously in the literature (see e.g. \cite{acta,micha}). Not to dwell on this once again, we just give the Table~\ref{tab:alapdg}
which summarises extracted properties of the putative anti-decuplet nucleon resonance and relevant references.  
{\small
\begin{table}[h]
  \begin{center}
    \begin{tabular}{lccc}  
    observable   &     extracted value & \     refs. (neutron data) &\ refs. (proton data)  \\
    \hline
      mass (MeV)      &$1680\pm 15$  & \cite{gra0,kas,kru,wert,Compton}\cite{arndt}$^{\star)}$& \cite{acta,jetp,KPT,BG}  \cite{arndt}$^{\star)}$\\
     $\Gamma_{\rm tot}$ (MeV)  & $\le 40$     &    \cite{gra0,kas,kru,wert,Compton}\cite{arndt}$^{\star)}$& \cite{acta,jetp,KPT,BG} \cite{arndt}$^{\star)}$ \\
      $\Gamma_{\pi N}$ (MeV) &    $\le 0.5$ &      \cite{arndt}$^{\star)}$&  \cite{arndt}$^{\star)}$ \\
      $\sqrt{{\rm Br}_{\eta N}} |A_{1/2}^n| \ (10^{-3}\ {\rm GeV}^{-1/2})$   & 12-18     &      \cite{az,wert} & \\
$\sqrt{{\rm Br}_{\eta N}} |A_{1/2}^p| \ (10^{-3}\ {\rm GeV}^{-1/2})$         & 1-3         &                                  &   \cite{acta,jetp,KPT,BG}  \\
    \end{tabular}
    \caption{ {\small Our estimate of properties of the putative narrow N$^*$ extracted from the data.  
    $^{\star)}$In Ref.~\cite{arndt} the elastic $\pi N$ scattering data were analyzed
    and the tolerance limits for N$^*$ parameters were obtained. The preferable  quantum numbers in this analysis are $P_{11}$.}}
    \label{tab:alapdg}
  \end{center}
\end{table} }
%


\end{document}